\documentclass[11pt]{article}
\usepackage{amsfonts}
\usepackage{amsmath}
\usepackage{amssymb}
\usepackage[english]{babel}
\usepackage{color}
\usepackage{geometry}
\usepackage{graphics}
\usepackage{graphicx}
\usepackage[colorlinks=true,urlcolor=blue,linkcolor=blue,citecolor=blue]{hyperref}
\usepackage{multicol}
\usepackage{natbib}
\usepackage{times}
\usepackage{lscape}

\renewcommand{\baselinestretch}{1.5}

\newcommand{\eg}{\textit{e.g.}}

\newcommand{\vbeta}{\text{\boldmath{$\beta$}}}
\newcommand{\vgamma}{\text{\boldmath{$\gamma$}}}

\newcommand{\mSigma}{\text{\boldmath{$\Sigma$}}}
\newcommand{\mOmega}{\text{\boldmath{$\Omega$}}}

\newcommand{\mX}{\mathbf{X}}

\newcommand{\vx}{\mathbf{x}}
\newcommand{\vh}{\mathbf{h}}
\newcommand{\vy}{\mathbf{y}}

\newcommand{\N}{\mathcal{N}}
\newcommand{\B}{\mathcal{B}}
\newcommand{\IG}{\mathcal{IG}}

\newcommand{\minus}{\text{\boldmath{$-$}}}
\newcommand{\plus}{\text{\boldmath{$+$}}}
\newcommand{\cross}{\text{$\times$}}
\renewcommand{\circle}{\text{$\bigcirc$}}
\renewcommand{\triangle}{\text{$\bigtriangleup$}}

\title{\textbf{
	Is Bitcoin really a currency?\\
	A viewpoint of a stochastic volatility model
	\thanks{
		We would like to thank Editage (\href{https://www.editage.com}{www.editage.com}) for English language editing.
		We would also like to thank the editor of this journal and the reviewer for their valuable comments and suggestions, which improve the paper substantially.
		This work is partially supported by JSPS KAKENHI (\#20H00080 and \#20K01590).
		Any views expressed in this paper represent those of the authors only and not necessarily of the PwC Consulting LLC.
	}
}}
\author{
	Noriyuki Kunimoto
	\thanks{
		PwC Consulting LLC.
		Email: \texttt{\href{mailto:kunimotonoriyuki@hotmail.co.jp}{kunimotonoriyuki@hotmail.co.jp}}
	}
	\and
	Kazuhiko Kakamu
	\thanks{
		Corresponding author.
		Graduate School of Economics, Nagoya City University, Yamanohata 1, Mizuho-cho, Mizuho-ku, Nagoya 467-8501, Japan.
		Email: \texttt{\href{mailto:kakamu@econ.nagoya-cu.ac.jp}{kakamu@econ.nagoya-cu.ac.jp}}
	}
}
\date{}
\begin{document}
\maketitle
\begin{abstract}
	Using the asymmetric stochastic volatility model, this study investigates the day-of-the-week and holiday effects on the returns and volatility of Bitcoin from January 1, 2013 to August 31, 2019; in this context, we also discuss the characteristics of Bitcoin as a financial asset.
	The results of the estimation are threefold.
	First, the finding shows a small day-of-the week effect in volatility on Saturday and Sunday than in the rest of the week.
	Second, although the holiday effects are examined in active trading countries, namely Japan, China, Germany, and the United States, the positive post-holiday effect on the returns and weak positive pre-holiday effect on the volatility are only observed in the United States.
	Finally, the asymmetry effect is not observed.
	A comparison of Bitcoin to several assets such as stock, currency, and gold shows Bitcoin's positioning between stock, currency, and gold in relation to the week and holiday effects, its reaction to federal funds and medium of exchange characteristics, and the lack of asymmetry effect.
	\vspace*{8pt}


\end{abstract}

\clearpage

\section{Introduction}

Bitcoin has been the most popular cryptocurrency since its launch by \cite{SN08}. The reasons behind Bitcoin's popularity may be attributed to some of the following features. First, it can be traded throughout the day and year and has low transaction costs, unlike traditional financial assets.
Second, Bitcoin is a peer-to-peer system of electronic cash that facilitates online currency exchange between two parties without the intervention of financial intermediaries.
Since Bitcoin is not supported by any central bank or government and it can be traded without a regulatory framework, Bitcoin uses the proof of work system in order to prevent illegal transactions; this system accurately combines transaction data on cryptocurrency with blockchain.
Although the proof of work facilitates correct and secure transactions on the peer-to-peer network without a control center, Bitcoin has been stolen by hackers several times.
Some of them seem to relate to events as the price pop up or down amazingly.
While there was a stable increase in its price until 2012, Bitcoin surged to \$1,150 in late November and, subsequently, declined below \$500 by mid-December in the same year.
The Bitcoin prices have gained significantly after the price pop in 2013. However, the prices plummeted after federal authorities seized the Silk Road website. This was followed by another massive price decline when Mt. Gox (a popular Bitcoin exchange) halted withdrawals due to technical issues. After an investigation, Mt. Gox revealed that it had lost 850,000 Bitcoins because transaction malleability.
This event caused Bitcoin prices to fall from \$867 to \$439 while raising questions over the security of Bitcoin and disturbing the liquidity in the currency.
There was a low increase in prices for up 2 years, which rose again until they reached around \$19,000 and, subsequently, crashed in December 2017.
Unlike a stable fiat currency, since the prices of Bitcoin fluctuate sharply, investors and financial experts tend to regard Bitcoin as a speculative asset.

There is ambiguity regarding whether Bitcoin should be categorized as currency.
Moreover, there is another question whether Bitcoin is different from the financial assets like stock and gold.
We focus on the characteristics on currency and the other assets; the day-of-the-week and holiday effects and asymmetry, which are (not) observed in the previous literature.
By using the asymmetric stochastic volatility (SV) model, this study aims to reveal the day-of-the-week and holiday effects on Bitcoin's returns and volatility and present where Bitcoin is placed in financial assets.
In this regard, this study redefines the holiday effect to differentiate it from the calendar hypothesis by \cite{KF80}, as per which returns are generated continuously in calendar time.
If a market is closed, returns on the day of opening are higher than those on other days.
However, Bitcoin transactions are conducted on a holiday, and hence the conventional definition for the holiday effect is not valid.
Therefore, in the context of this study, a holiday effect means that a return is not only different on the day before or after a certain holiday but also on a certain holiday.
According to a paper published on April 29, 2019 by DataLight, a company that provides in-dept analysis of cryptocurrencies, Japan, China, Germany, and the United States are relatively active trading countries. Hence, this study's model incorporates the pre-holidays, holidays, and post-holidays in these countries. 

The contribution of our study is threefold.
First, to the best of our knowledge, this is the first study to observe a small day-of-the-week effect in volatilities on Saturday and Sunday.
Second, unlike previous studies, this is the first study to examine the holiday effect on Bitcoin's returns and volatility, and the United States is the only country for which we observe the positive post-holiday effect on returns and the weak positive pre-holiday effect on the volatility.
Finally, an asymmetry effect is not observed in the Bitcoin market, even after employing a different model from the previous literature.

The rest of this paper is organized as follows.
In Section \ref{sec:previous}, we briefly explain the previous literature.
In Section \ref{sec:model}, we introduce our model, which estimates Bitcoin's return and volatility.
Section \ref{sec:post} provides details of the posterior analysis.
Section \ref{sec:data} presents the dataset and conducts preliminary analyses, such as summarizing the descriptive statistics.
Section \ref{sec:emp} provides the empirical results and, on the basis of an examination of several assets and the study's findings, suggests the type of asset category to which Bitcoin belongs.
The final section provides the conclusion.

\section{Previous literature}\label{sec:previous}
In this section, we briefly review the previous literature.
We mainly focus on the empirical characteristics of the asymmetry and the day-of-the-week and holiday effects and the models, which are utilized to examine the effects on currency and financial assets.

According to \cite{YO08}, a currency does not generate an asymmetric effect; in this phenomenon, the volatility of traditional financial assets tends to increase more after a negative shock than after a positive shock of the same magnitude, although \cite{MM02} and \cite{MM07} observe it. \cite{JB16} and \cite{SK18} discover the asymmetric effect in the Bitcoin market, whereas \cite{AHD16} do not observe this effect, suggesting that Bitcoin may be categorized as other financial assets such as stock and gold.

Besides the above phenomenon, the literature has also studies the day-of-the-week and holiday effects in the traditional financial markets, especially in stock markets.
The day-of-the-week effect means that a return is different on a certain day of the week.
The holiday effect means that a return is different on a day before or after a certain holiday. 
These are considered calendar effects or anomalies.
\cite{IT06} reveals the day-of-the week and holiday effects on the returns and volatility in the US stock markets.
Studies have also examined these calendar effects on the returns and volatility in the stock markets in Japan \citep{FP96,HT04}, Europe \citep{OD12,JC13,RI15}, and China \citep{TR14}.
The day-of-the-week effects on the returns and volatility are also reported in the gold markets \citep{BL06}; however, limited research has been done on holiday effects on the returns and volatility in these markets.
Recent studies have also not examined the existence of calendar effects in currency markets \citep{DH88,HB07,SK18}.
In this context, it must be noted that calendar effects refute the efficient market hypothesis (EMH) by \cite{EF65}. This hypothesis states that asset prices reflect all the available information, and hence no analysis can help investors to acquire abnormal returns and predict the future value.
Several studies do not support the EMH for the Bitcoin market.
Using a Hurst exponent analysis of daily data, \cite{JY18} find a strong persistence in the returns series, and hence conclude that the Bitcoin market is inefficient.
This conclusion is supported by \cite{DA19} because of the unsuitability of the random walk model.
The evidence that Bitcoin market is inconsistent with the EMH suggests the existence of some calendar effects.
\cite{GM19} examine the day-of-the-week effect in several cryptocurrency markets and only find the positive Monday effect on Bitcoin returns.
\cite{DM19b} reveal that the exchange level determines the patterns of the day-of-the-week effects on Bitcoin returns.
They find positive Monday and Thursday effects on Bitcoin returns when it is priced in USD and JPY.
However, \cite{DB19} do not observe persistent anomalies in the day-of-the week effects on Bitcoin returns, but find the trading volume on Saturday and Sunday smaller than that on the weekdays.

The main models used to estimate the returns and volatility together are the generalized autoregressive conditional heteroskedastic (GARCH) models, proposed by \cite{TB86}, and the stochastic volatility (SV) models \citep[\eg,][]{SK98}.
The former models the conditional variance as a deterministic function of the past data, while the latter models it as a latent variable.
\cite{DY18} find the day-of-the week effects on the returns and volatility of Bitcoin for three subsamples, from July 2010 to October 2017, using the GARCH model with the maximum likelihood estimation.
They state that Mondays are generally associated with higher returns and volatility.
By comparing GARCH with SV models, \cite{AK19} determine which models are better for describing the dynamics of Bitcoin.
They conclude that SV models outperform the GARCH models in majority of the cases because SV models are more robust to misspecification and radical changes.
Using an asymmetric SV model, \cite{DM19a} show that the day-of-the-week effects on returns vary with the sample period, and higher volatilities are observed on Monday and Thursday.

In this paper, we estimate the asymmetric SV model incorporating the day-of-the-week and holiday effects.
In this sense, the approach is close to \cite{DM19a}.
However, our approach is different from their approach in the following points.
Their model consists of a constant and all days as dummy variables.
In this specification, it is difficult to differentiate one day-of-the-week effect from the other.
Therefore, we specify the model including a constant term as one of the regressors.
Besides, their model includes the asymmetry effect as a dummy variable, which is not the general asymmetric SV model at present.
Our approach includes the asymmetry effect as a parameter $\rho$ in the model.
As far as we know, the holiday effects in Bitcoin market using a GARCH or an SV model have not been examined.
In the next section, we introduce an asymmetric SV model, which incorporates the day-of-the-week and holiday effects both in returns and volatilities mentioned above.

\section{The model}\label{sec:model}

This study uses the asymmetric SV (ASV) model \citep[\eg~][]{JN09,JN12} to investigate Bitcoin's returns and volatility.
The ASV models are used to model a changing variance of time series under flexible restrictions.
Before explaining the ASV model, we discuss the standard SV model. 
Let $y_{t}$ and $h_{t}$ denote returns and unobserved log-volatility.
Then, the standard SV model is given by
\begin{align}
	y_{t} &= \exp\left( \frac{h_{t}}{2} \right) \epsilon_{t},\quad t = 1, \cdots, T,\label{eqn:observation}\\
	h_{t+1} &= \mu + \phi(h_{t} - \mu ) + \eta_{t},\quad t = 2, \cdots, T-1,
	\label{eqn:state}
\end{align}
where $\epsilon_{t} \sim \N(0,1)$ and $\eta_{t} \sim \N(0, \sigma^{2})$, and $\epsilon_{t}$ and $\eta_{t}$ are independent.
For the initial distribution $h_{1}$, we assume
\begin{align}
	h_{1} \sim \N \left( \mu, \frac{\sigma^{2}}{1-\phi^{2}}\right),
	\label{eqn:init}
\end{align}
$|\phi| < 1$ is assumed for the stationary condition. This parameter denotes the intensity of persistence in the volatility, that is, the closer the $\phi$ is to 1, the higher will be its persistence.

In the ASV model, $\epsilon_{t}$ is correlated with $\eta_{t}$.
These correlations can be considered on the basis of two patterns: the model that correlates $\epsilon_{t}$ with $\eta_{t-1}$ and the model that correlates $\epsilon_{t}$ with $\eta_{t}$.
\cite{JY05} argues that the former model has two main drawbacks.
One is that a martingale difference sequence is not held, and hence it is inconsistent with the EMH, and the other is that it is difficult to interpret the asymmetry effect because the correlation parameter has two effects of asymmetry and skewness.
However, the latter model proves to overcome these drawbacks.
Therefore, we use the model that correlates $\epsilon_{t}$ with $\eta_{t}$ and we assume as follows:
\begin{align}
	\left(
		\begin{array}{c}
			\epsilon_{t}\\
			\eta_{t}
		\end{array}
	\right) \sim \N \left( \mathbf{0}, \mSigma \right), \quad
	\mSigma =
	\left(
		\begin{array}{cc}
			1 & \rho\sigma\\
			\rho\sigma & \sigma^{2}
		\end{array}
	\right).
	\label{eqn:errors}
\end{align}
The asymmetry effect presented by the correlation coefficient $\rho$ is expected to be negative.
The correlation coefficient $\rho = 0$ becomes the standard SV model.

To examine the day-of-the-week and holiday effects, two sets of regressors are incorporated into the observation equation \eqref{eqn:observation} and state-space equation \eqref{eqn:state}.
Therefore, we consider the following ASV model given by
\begin{align}
	y_{t} &= \vx_{t}^{\prime}\vbeta + \exp{ \left(\frac{h_{t}}{2}\right)} \epsilon_{t},\quad t = 1, \cdots, T,\label{eqn:observation2}\\
	h_{t+1} &= \vx_{t+1}^{\prime}\vgamma+ \phi( h_{t} - \vx_{t}^{\prime} \vgamma ) + \eta_{t},\quad t = 2, \cdots, T,\label{eqn:state2}
\end{align}
and
\begin{align}
	h_{1} \sim \N \left( \vx_{1}^{\prime} \vgamma, \frac{\sigma^{2}}{1-\phi^{2}}\right),
	\label{eqn:init2}
\end{align}
where $\vx_{t}$ is a $k \times 1$ vector of covariates, which include the constant term in the first element, and $\vbeta$ and $\vgamma$ are $k \times 1$ vectors of parameters for returns and volatilities, respectively.
For $\epsilon_{t}$ and $\eta_{t}$, we assume the same distribution with \eqref{eqn:errors}.
Finally, it should be mentioned that the length of the time period in state-space equation \eqref{eqn:state2} is changed from $T-1$ in \eqref{eqn:state} to $T$.
Following this change, the Markov chain Monte Carlo (MCMC) algorithm should be modified appropriately.

\section{Posterior analysis}\label{sec:post}

It is difficult to estimate even the parameters of the standard SV model by maximum likelihood because their likelihood functions are not easily tractable.
Moreover, our model considers a correlation between returns and volatilities and two sets of regressors in order to examine the asymmetry effect and day-of-the week and holiday effects. Therefore, it is more difficult to estimate the parameters of the model.
Although it is possible to evaluate the likelihood for a given set of parameters using a simulation-based method, the computational burden would increase with every iteration of the filtering procedure.
Alternatively, we employ a Bayesian MCMC method to overcome the computational burden.
This approach provides a precise and efficient estimation of the SV models.
Therefore, in this section, we use the MCMC procedure to estimate the parameters of the model.

Let $\vy = (y_{1}, y_{2},\ldots,y_{T})^{\prime}$, $\vh = (h_{1}, h_{2}, \ldots, h_{T+1})^{\prime}$, and $\mX = (\vx_{1}, \vx_{2},\ldots, \vx_{T+1})^{\prime}$, respectively.
If the likelihood function for \eqref{eqn:errors} -- \eqref{eqn:init2} is defined by $L(\vy | \vbeta, \vgamma, \phi, \rho, \sigma^2, \vh, \mX)$ and the prior distributions for $\vbeta$, $\vgamma$, $\phi$, $\rho$, and $\sigma^{2}$ are set as $\pi(\vbeta)$, $\pi(\vgamma)$, $\pi(\phi)$, $\pi(\rho)$, and $\pi(\sigma^2)$, respectively, the joint posterior distribution of the model is expressed by
\begin{align}
	\pi(\vbeta, \vgamma, \phi, \rho, \sigma^{2}, \vh | \vy, \mX) \propto \pi(\vbeta) \pi(\vgamma) \pi(\phi) \pi(\rho) \pi(\sigma^{2}) L(\vy | \vbeta, \vgamma, \phi, \rho, \sigma^2, \vh, \mX).\label{eqn:posterior}
\end{align}
From the joint posterior distribution, if the full conditional distribution for each parameter is derived from \eqref{eqn:posterior}, we can generate MCMC samples as follows:

\begin{enumerate}
	\item Initialize $\vbeta$, $\vgamma$, $\phi$, $\rho$, $\sigma^{2}$ and $\vh$.
	\item Sample $\vbeta | \vgamma, \phi, \rho, \sigma^{2}, \vh, \vy, \mX$.\label{Step2}
	\item Sample $\vgamma | \vbeta, \phi, \rho, \sigma^{2}, \vh, \vy, \mX$.
	\item Sample $(\phi, \rho, \sigma^{2})$ by
		\begin{enumerate}
			\item $\phi | \vbeta, \vgamma, \rho, \sigma^{2}, \vh, \vy, \mX$,
			\item $\rho | \vbeta, \vgamma, \phi, \sigma^{2}, \vh, \vy, \mX$.
			\item $\sigma^{2} | \vbeta, \vgamma, \phi, \rho, \vh, \vy, \mX$,
		\end{enumerate}
	\item Sample $\vh | \vbeta, \vgamma, \phi, \rho, \sigma^2, \vy, \mX$.
	\item Go to \ref{Step2}. until converge.
\end{enumerate}

Finally, to conduct Bayesian analysis, we assume that
\begin{align*}
	\vbeta \sim \N(\vbeta_{0}, \mSigma_{0}),\quad \vgamma \sim \N(\vgamma_{0}, \mOmega_{0}),\quad \frac{\phi + 1}{2} \sim \B(\alpha_{0},\beta_{0}),\quad \frac{\rho + 1}{2} \sim \B(\delta_{0},\zeta_{0}),\quad \sigma^{2} \sim \IG(\nu_{0}/2, \lambda_{0}/2),
\end{align*}
for prior distributions, where $\B$ and $\IG$ denote beta and inverse gamma distributions, respectively.
In the following subsections, we provide a brief description of each full conditional distribution.

\subsection{Sampling \texorpdfstring{\vbeta}{beta} and \texorpdfstring{\vgamma}{gamma}}
The full conditional posterior distribution for $\displaystyle \vbeta$ is $\vbeta | \vgamma, \phi, \rho, \sigma^{2}, \vh, \vy, \mX \sim \N( \hat{\vbeta}, \hat{\mSigma})$, where
\begin{align*}
	\hat{\mSigma} &= \left( \sum_{t=1}^{T}\frac{\vx_{t} \vx_{t}^{\prime}}{(1-\rho^{2})\exp(h_{t})} + \mSigma_{0}^{-1} \right)^{-1},\\
	\hat{\vbeta} &= \hat{\mSigma} \left( \sum_{t=1}^{T}\frac{\vx_{t}\left( y_{t} - \rho \sigma^{-1} \exp\left(\frac{h_{t}}{2}\right)\left(h_{t+1} - \vx_{t+1}^{\prime}\vgamma - \phi(h_{t} - \vx_{t}^{\prime} \vgamma)\right) \right)}{(1 - \rho^{2})\exp{(h_{t})}} + \mSigma_{0}^{-1} \vbeta_{0} \right),
\end{align*}
and the full conditional distribution for $\vgamma$ is $\vgamma | \vbeta, \phi, \rho, \sigma^{2}, \vh, \vy, \mX \sim \N ( \hat{\vgamma}, \hat{\mOmega})$, where
\begin{align*}
	\hat{\mOmega} &= \left( \frac{(1-\phi^{2})\vx_{1} \vx_{1}^{\prime}}{\sigma^{2}} + \sum_{t=1}^{T}\frac{(\vx_{t+1} - \phi \vx_{t}) (\vx_{t+1} - \phi \vx_{t})^{\prime}}{(1-\rho^{2})\sigma^{2}} + \mOmega_{0}^{-1} \right)^{-1},\\
	\hat{\vgamma} &=  \hat{\mOmega} \left( \frac{(1-\phi^{2})\vx_{1}h_{1}}{\sigma^2} + \sum_{t=1}^{T}\frac{(\vx_{t+1} - \phi \vx_{t}) \left( h_{t+1} - \phi h_{t} - \rho \sigma \exp\left( -\frac{h_{t}}{2}\right)(y_{t} - \vx_{t}^{\prime} \vbeta )\right)}{(1-\rho^{2})\sigma^{2}} + \mOmega_{0}^{-1}\vgamma_{0}\right).
\end{align*}
These parameters are easily sampled by Gibbs sampler \citep[see][]{GS90}. 

\subsection{Sampling \texorpdfstring{$\phi$}{phi}, \texorpdfstring{$\rho$}{rho} and \texorpdfstring{$\sigma^{2}$}{sigma2}}
The full conditional posterior distribution for $\phi$, $\rho$ and $\sigma^{2}$ is given by
\begin{align}
	\lefteqn{\pi(\phi, \rho, \sigma^{2} | \vbeta, \vgamma, \vh, \vy, \mX)}\nonumber\\
	& \propto \left(\frac{1+\phi}{2}\right)^{\alpha_{0} - 1} \left( \frac{1 - \phi}{2} \right)^{\beta_{0} - 1} \left( \frac{1+\rho}{2}\right)^{\delta_{0} - 1} \left(\frac{1-\rho}{2}\right)^{\zeta_{0} - 1} \left(\sigma^{2}\right)^{-\left(\frac{\nu_{0}}{2} + 1\right)} \exp\left( - \frac{\lambda_{0}}{2\sigma^{2}} \right) \nonumber\\
	& \times (1-\phi^{2})^{\frac{1}{2}} \left(\sigma^{2}\right)^{-\left(\frac{T + 1}{2}\right)} (1-\rho^{2})^{\frac{T}{2}} \exp\left( - \frac{( 1 - \phi^{2})( h_{1} - \vx_{1}^{\prime} \vgamma)^{2}}{2\sigma^{2}} \right)\nonumber\\
	& \times \exp\left( - \frac{1}{2}\sum_{t=1}^{T}\left( \frac{\left( h_{t+1} - \vx_{t+1}^{\prime}\vgamma - \phi( h_{t} - \vx_{t}^{\prime}\vgamma) - \rho \sigma \exp\left(-\frac{h_{t}}{2}\right)(y_{t}-\vx_{t}^{\prime} \vbeta) \right)^{2}}{(1-\rho^{2})\sigma^{2}} \right) \right).
	\label{fcd:MH}
\end{align}
As it is difficult to sample from the standard distribution, we sample each parameter using the random walk Metropolis-Hastings (MH) algorithm. 
The suitable choice of step sizes used in the random walk chain is determined by the procedure in \cite{GH02}. 

\subsection{Sampling \texorpdfstring{$\vh$}{h}}
The full conditional distributions of $h_{t}$ for $t=1, 2, \ldots, T + 1$ are different depending on $t$.
If $t = 1$, then the full conditional distribution of $h_{1}$ is given as follows:
\begin{align*}
	\lefteqn{\pi(h_{1} | \vbeta, \vgamma, \phi, \rho, \sigma^{2}, h_{2}, y_{1}, \vx_{1}, \vx_{2})}\\
	& \propto \exp\left( -\frac{h_{1}}{2} \right) \exp\left( - \frac{(y_{1} - \vx_{1}^{\prime} \vbeta )^{2}}{2\exp({h_{1})}} \right) \exp\left( -\frac{(1-\phi^{2})(h_{1}-\vx_{1}^{\prime} \vgamma)^{2}}{2\sigma^{2}} \right)\\
	& \times \exp\left( - \frac{\left( h_{2} - \vx_{2}^{\prime}\vgamma - \phi(h_{1}-\vx_{1}^{\prime}\vgamma) - \rho \sigma \exp\left(-\frac{h_{1}}{2}\right)(y_{1}-\vx_{1}^{\prime} \vbeta)\right)^{2}}{2 (1-\rho^{2}) \sigma^{2}} \right).
\end{align*}
If $t = 2, \ldots, T$, the full conditional distribution of $h_{t}$ is given as follows:
\begin{align*}
	\lefteqn{\pi(h_{t} | \vbeta, \vgamma, \phi, \rho, \sigma^{2}, h_{t-1}, h_{t}, h_{t+1}, y_{t-1}, y_{t}, \vx_{t-1}, \vx_{t}, \vx_{t+1})}\\
	& \propto \exp\left( -\frac{h_{t}}{2} \right) \exp\left( -\frac{(y_{t}-\vx_{t}^{\prime} \vbeta)^{2}}{2\exp(h_{t})} \right)\\
	& \times \exp\left( -\frac{\left( h_{t+1}-\vx_{t+1}^{\prime} \vgamma - \phi(h_{t}-\vx_{t}^{\prime}\vgamma)-\rho\sigma\exp\left(-\frac{h_{t}}{2}\right)(y_{t}-\vx_{t}^{\prime}\vbeta)\right)^2}{2 (1-\rho^{2}) \sigma^{2}} \right)\\
	& \times \exp\left( -\frac{\left( h_{t}-\vx_{t}^{\prime} \vgamma-\phi(h_{t-1}-\vx_{t-1}^{\prime} \vgamma)-\rho\sigma\exp\left(-\frac{h_{t-1}}{2}\right)(y_{t-1}-\vx_{t-1}^{\prime} \vbeta) \right)^2}{2 (1-\rho^{2}) \sigma^{2}} \right).
\end{align*}
If $t = T + 1$, the full conditional distribution for $h_{T+1}$ is as follows:
\begin{align*}
	h_{T+1} | \vbeta, \vgamma, \phi, \rho, \sigma^2, h_{T}, y_{T}, \vx_{T} \sim \N\left( \mu_{T+1}, \sigma^{2}( 1- \rho^{2} )\right),
\end{align*}
where $\mu_{T+1} = \vx_{T+1}^{\prime}\vgamma + \phi( h_{T} - \vx_{T}^{\prime}\vgamma) + \rho\sigma\exp\left(-\frac{h_{T}}{2}\right)(y_{T} - \vx_{T}^{\prime}\vbeta)$.

In the case of $t = T+1$, it is easily drawn from the normal distribution.
However, it is difficult to sample from the standard distributions, for $t = 1, 2, \ldots, T$.
Therefore, the random walk MH algorithm is applied similarly as $\phi$, $\rho$, and $\sigma^{2}$.
The suitable choice of step sizes by \cite{GH02} is also applied.
This method is applied to the SV models by \cite{NKO12,NK15}.

\section{Data and descriptive statistics}\label{sec:data}

For the Bitcoin Price Index (BPI), we use daily prices calculated in USD, from January 1, 2013 to August 31, 2019; the data are collected from an open source crypto-asset analytics database, \href{https://coinmetrics.io}{Coin Metrics} according to \cite{OY19}.
Daily prices are based on the closing prices at 00\,:\,00 UTC.
Using daily data, it is difficult to distinguish holiday effects clearly in each country on the basis of time zones.
Since we consider the day-end closing prices of Bitcoin traded in each country, the difference in time zones may not significantly affect our findings; this is because, at 00\,:\,00 UTC, it is midnight in Germany and the United States and early morning in Japan and China.
Hence, we do not consider the difference in time zones.
Daily return is defined as $y_{t} = (\log{P_{t}} - \log{P_{t-1}}) \times 100$ where $P_{t}$ denotes the closing price on date $t$.

\begin{figure}[tbh]0
	\includegraphics[width = \textwidth, bb = 0 0 576 360]{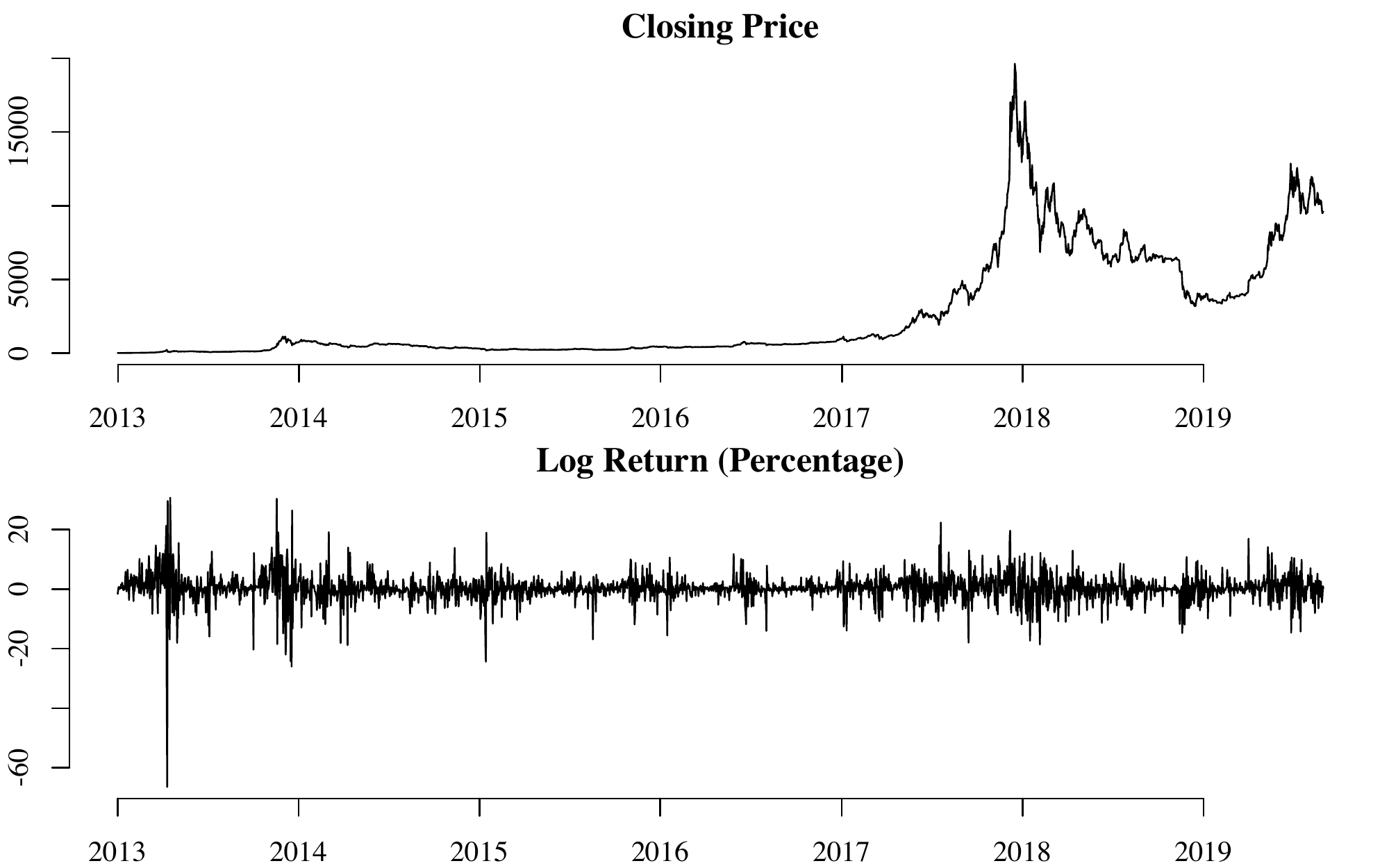}
	\caption{Daily data of Bitcoin Price Index based on Coin Metrics from January 1, 2013 to August 31, 2019 }
	\label{fig1}
\end{figure}

Figure \ref{fig1} shows the closing prices and daily returns in the upper and lower plots, respectively.
The upper plot presents the evolution of the Bitcoin prices from January 1, 2013 to August 31, 2019.
The closing prices of Bitcoin have seen a significant increases and decreases.
In the crash of December 17, 2017, the closing price of Bitcoin reached an all-time high of \$$19,640.5$ before crashing.
This price continued to fall through 2018, plummeting to \$$3,687.2$ on December 31, 2018.
However, the price rose again in January 2019, before rising to \$$12,863.5$ in June 26, 2019.
In the lower plot, the highest daily return is larger than $25$\%, while the lowest daily return is smaller than $25$\%, except for early April 2013.
The series of returns suggest that the prices of Bitcoin have volatile movements.

Table \ref{tab1} summarizes the descriptive statistics for daily returns and returns on the day-of-the-week.
The sample comprises 2434 observations.
``Mean'' stands for the means, and ``SD'' represents the standard deviation of the sample.
``Min,'' ``Max,'' ``Skew,'' and ``Kurt'' denote minimum, maximum, skewness, and kurtosis, respectively.
The largest and smallest mean returns are observed on Monday and Wednesday, respectively.
The standard deviation is the largest on Thursday and the smallest on Sunday.
The returns on Bitcoin are lowest and highest on Thursday and Wednesday, respectively.
The skewness is positive on Monday and Friday, while it is negative on the other weekdays.
The kurtosis is greater than zero.

To examine the holiday effects, we create the dummy variables of the pre-holidays, holidays, and post-holidays in each country; we refer to \href{https://www.timeanddate.com}{timeanddate.com}, by following \cite{OD12}.
Since Saturday and Sunday are treated as part of the days of the week, we do not include Saturday and Sunday in pre-holidays, holidays, and post-holidays of each country, except when they overlap with national and religious holidays.
Table \ref{tab2} summarizes the descriptive statistics for returns in the pre-holidays, holidays, and post-holidays in Japan, China, Germany, and the United States.
The largest and smallest mean returns are observed in the post- and pre-holidays, respectively, in the United States and Germany.
The standard deviation is the largest and smallest in the post-holidays and holidays, respectively, in the United States.
The returns on Bitcoin are lowest in the pre-holidays and holidays, respectively, in Germany and China.
However, the returns are the highest in the holidays in Japan.
The skewness is positive in the pre-holidays and holidays in Japan and in the post-holidays in Germany, while it is negative in other holidays.
The kurtosis is larger than zero in the pre-holidays, holidays, and post-holidays in all countries.

\section{Empirical analysis}\label{sec:emp}

For the Bayesian approach, we set the hyper-parameters for prior distribution.
For $\phi$, $\rho$ and $\sigma^{2}$, we assume $\alpha_{0} = 20.0$, $\beta_{0} = 1.5$, $\delta_{0} = 1.0$, $\zeta_{0} = 1.0$, $\nu_{0} = 5.0$, and $\lambda_{0} = 0.01$, respectively.
The mean vectors and covariance matrices for $\vbeta$ and $\vgamma$ are the zero vectors and diagonal matrices with 100, respectively.
We set $E(\phi) = 0.86$ and $\sqrt{Var(\phi)} = 0.11$ to reflect the high persistence of the volatility, as considered in the literature \citep[see \eg,][]{YO07,YO08,JN09}.
In the MCMC sampling of the posterior distribution, we generate $200,000$ samples, of which we discard $50,000$ samples as burn-in period.
With the remaining $150,000$ samples, we store $15,000$ thinned samples every 10th draw after the initial burn-in period.
These computational results were obtained using Ox version 8.02 (OS\_X\_64/U) \citep[see][]{D13}, and all the figures were drawn using the R version 4.0.2 \citep[see][]{R19}.

We compute the following two diagnostics.
First, to check the convergence of MCMC, we compute the convergence diagnostic (CD) given by \cite{GJ92}.
CD represents the p-value based on the test statistic on the difference between two sample means (i.e., dividing all the generated random draws into three parts, we compute two sample means from the first 20\% and last 50\% of the random draws); in this case, the test statistic is asymptotically distributed as a standard normal random variable.
We confirm that the random draws generated by MCMC do not converge to the random draws generated from the target distribution when CD is less than $0.01$.
Second, to check the efficiency of the MCMC algorithm, we estimate the inefficiency factor (IF).
IF is defined as $1 + 2 \sum_{s=1}^{\infty}\rho_{s}$ where $\rho_{s}$ is the sample autocorrelation at lag $s$ calculated from the sampled values.
It is used to measure how well the chain mixes, and it represents the ratio of the numerical variance of the sample posterior mean to the variance of the sample mean from the hypothetical uncorrelated draws \citep[\eg,][]{SC01}.

Finally, we explain the construction of $\vx_{t}$.
It is a vector of the day-of-the-week dummy variables, except for Wednesday and pre-holiday, holiday, and post-holiday dummy variables in Japan, China, Germany, and the United States in period $t$.
It means that we assign a constant to Wednesday because it lies in the middle of the week.
As a result, $\vx_{t}$ becomes $19 \times 1$ the vector of covariates.

Table \ref{tab3} and table \ref{tab4} exhibit estimation results of the observation equation and state-space equation in the SV model, respectively.
``Mean'' stands for the posterior means of parameters, while ``SD'' presents the standard deviation of the parameters.
The term ``95\%CI'' denotes 95\% credible intervals for posterior distributions.

\begin{figure}[tbh]
	\centering
	\includegraphics[width = \textwidth, bb = 0 0 504 504]{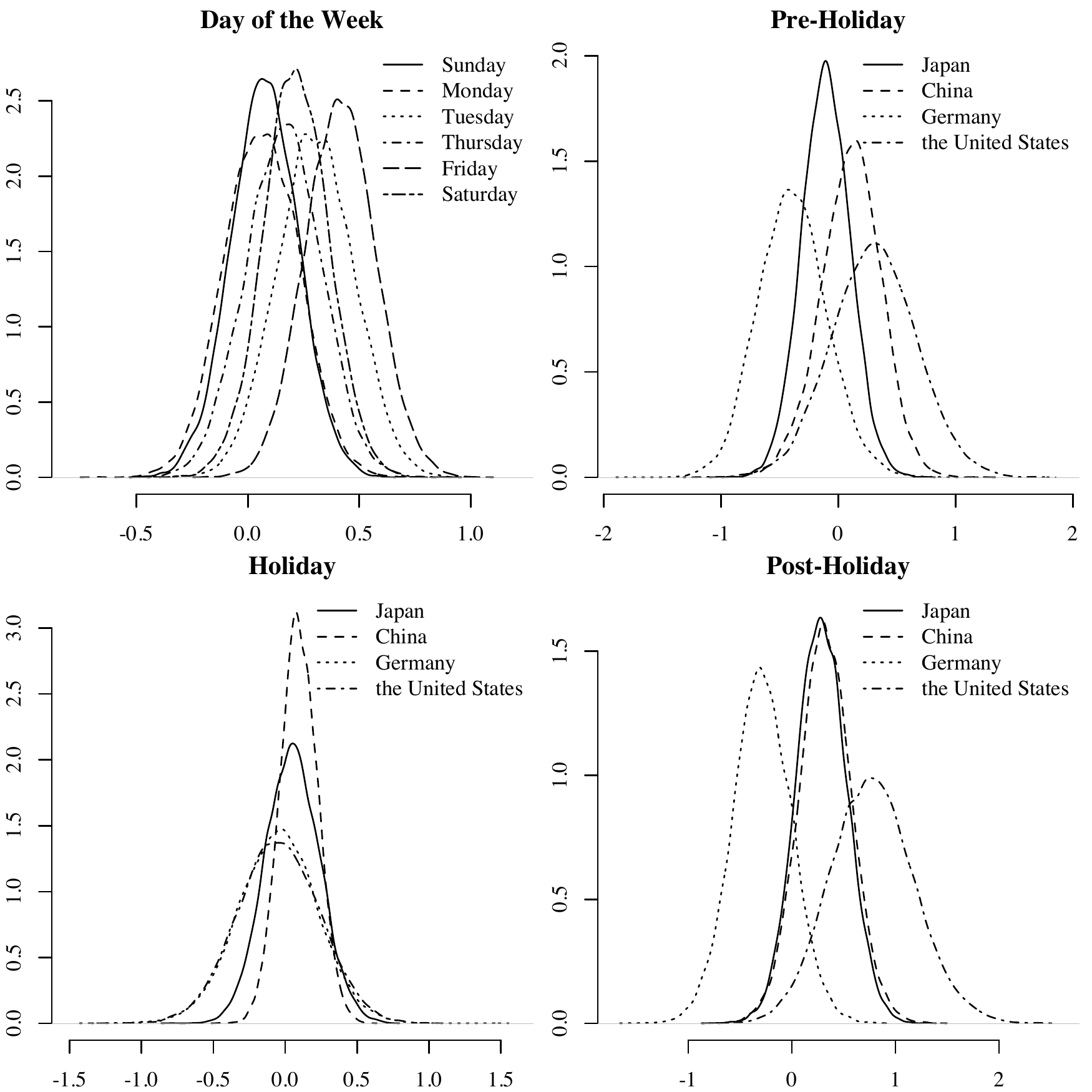}
	\caption{Marginal posterior densities of $\vbeta$}
	\label{fig2} 	
\end{figure}

First, we confirm the results of the observation equation according to Table \ref{tab3}.
Except for Friday, the posterior distributions include zeros in the 95\% credible intervals.
This result suggest that the return on Friday is larger than that on Wednesday.
With respect to holiday effects, the posterior distributions include zeros in the 95\% credible intervals, except for the post-holiday in the United States, suggesting that post-holiday returns in the United States are larger than those on Wednesday.
Therefore, Friday and the post-holiday in the United States make the return on that day larger.
Figure \ref{fig2} shows that the marginal posterior distributions of $\vbeta$ for the day-of-the-week, pre-holiday, holiday, and post-holiday effects on returns.
The posterior distributions on Friday and the post-holiday in the United States are positioned on the plus parts.

\begin{figure}[tbh]
	\centering	
	\includegraphics[width = \textwidth, bb = 0 0 504 504]{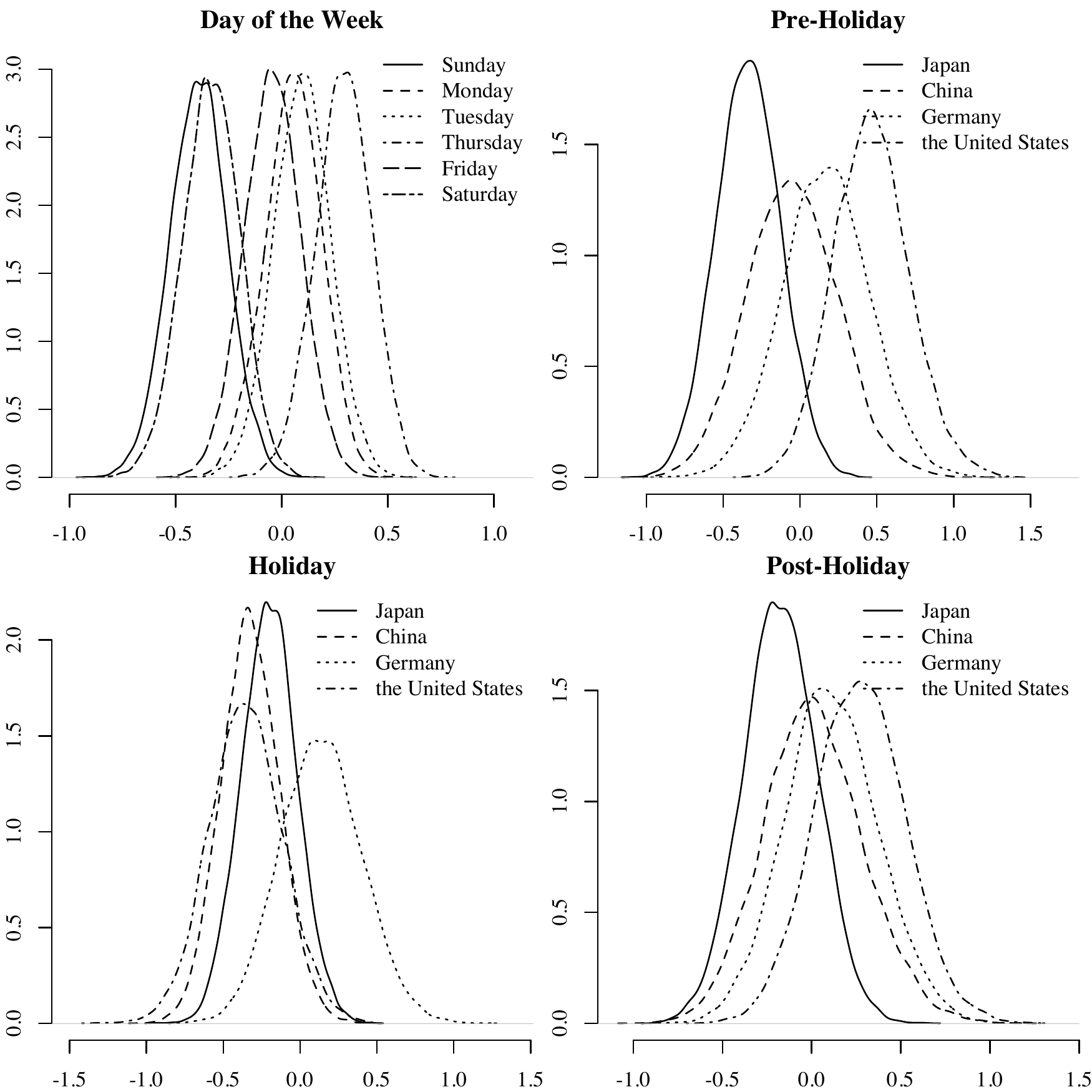}
	\caption{Marginal posterior densities of $\vgamma$}
	\label{fig3} 	
\end{figure}

Next, we confirm the results of the state-space equation according to Table \ref{tab4}.
The posterior mean of $\phi$ is 0.915, suggesting that the log volatility has high persistence.
The posterior mean of $\rho$ is positively 0.049; however, it might be zero because the 95\% credible interval includes zero, suggesting that there is no asymmetry effect in Bitcoin market.
Let us see the results of the day-of-the-week effects.
The credible intervals on Thursday, Saturday, and Sunday do not include zeros in the 95\% credible intervals.
This suggests a larger volatility on Thursday and a smaller volatility on Saturday and Sunday, compared to that on Wednesday.
In terms of holiday effects, zero is included in the 95\% credible interval for the pre-holiday in the United States.
However, if we change the credible interval from 95\% to 90\%, it does not include zero in the 90\% credible interval.
Moreover, the probability that is larger than zero is 0.97, suggesting that Bitcoin market is weakly subjected to the pre-holiday in the United States.
\footnote{
	We implement the estimation, which excludes the dummy variables of holidays.
	For pre-holiday effects in the United States, the result suggests that the estimated coefficient is different from the 95\% credible interval containing zero. 
}
Accordingly, Thursday and the pre-holiday in the United States amplify the following volatility, whereas Saturday and Sunday force them to be smaller.
Figure \ref{fig3} denotes the posterior distributions of $\vgamma$ for the day-of-the-week, pre-holiday, holiday, and post-holiday effects on the volatility.
The posterior densities on Thursday and the pre-holidays in the United States are placed on the plus parts.
However, the posterior densities on Saturday and Sunday are shown to be positioned on the minus parts.

The results are summarized as follows.
The day-of-the-week effects on Bitcoin's returns and volatility are not completely consistent with the literature.
Positive Tuesday and Friday effects on returns and positive Monday and Thursday effects on volatility are observed by \cite{DM19a}, whereas only positive Friday effect on returns and positive Thursday effect on the volatility are revealed in this study.
To the best of our knowledge, this is the first study to observe the negative Saturday and Sunday effects on the volatility.
Both estimated coefficients are smaller than those on Wednesday, suggesting that investors in the Bitcoin market may avoid trading it because of leisure on the weekends.
Next, the positive post-holiday effect on returns and the weak positive pre-holiday effect on volatility are only observed in the United States.
It is interesting to note that holiday effects are observed only in the United States America, despite the fact that Japan, China, and Germany are active trading countries; this finding suggests that the Bitcoin market is influenced by the behavior of investors subjected to the seasonal effect of national and religious holidays in the United States.
Although the method to incorporate the asymmetry effect into the model is different from that employed in the literature, no asymmetry effect is observed in this study, suggesting that there is no asymmetry effect in the Bitcoin market.

\begin{figure}[tbh]
	\includegraphics[width = \textwidth, bb = 0 0 576 360]{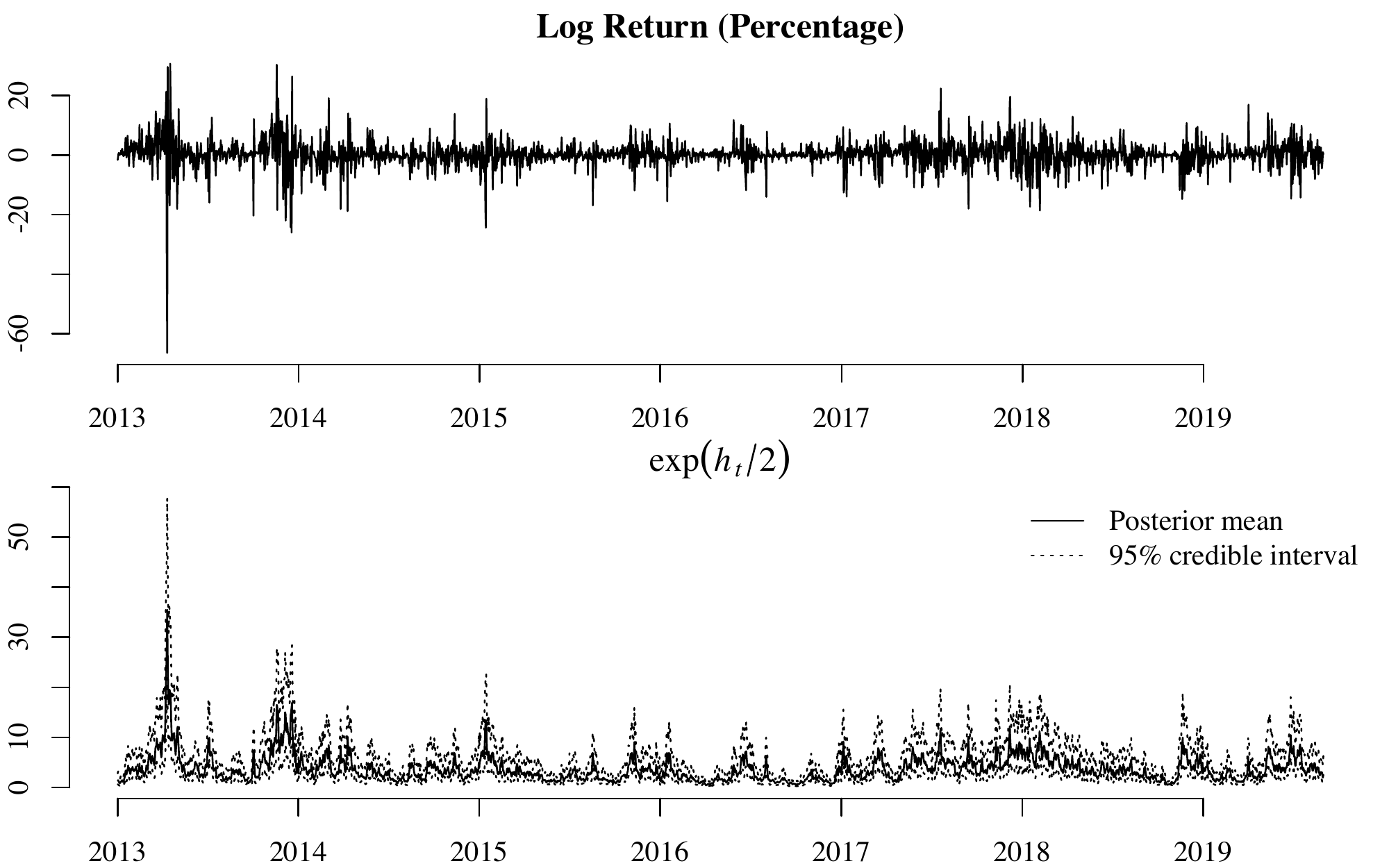}
	\caption{Daily data of Bitcoin Price Index based on Coin Metrics from January 1, 2013 to August 31, 2019 }
	\label{fig4}
\end{figure}

Finally, we examine the volatility.
The volatility represents a fluctuation in Bitcoin prices.
A sharp fluctuation in Bitcoin prices leads to either a significant gain or loss.
Hence, the volatility indicates a better index to avoid a risk.
Figure \ref{fig4} exhibits the posterior means and 95\% credible intervals of volatilities from the estimated results.
By comparing returns with estimated volatilities, we find that the estimated volatilities may capture the price fluctuations of Bitcoin.
The fluctuation of higher volatilities is presented from January 2013 to December 2013 than that in other periods.
After 2013, the volatilities repeatedly exhibit sharp up-down movements.

There is still ambiguity concerning the reason behind the occurrence of the day-of-the-week and holiday effects on the returns and volatility in the Bitcoin market.
In this regard, although several studies have conducted examinations on the calendar effects, they are not applicable to Bitcoin.
Bitcoin is similar to several traditional financial assets; however, all its characteristics are not completely consistent with those of a given asset.
Hence, based on the study's findings and the literature on stock, currency, and gold, we identify the assets that most closely reflect the characteristics of Bitcoin.

Table \ref{tab5} presents a summary of related literature.
While there are several studies on the day-of-the-week and holiday effects on stock markets, there are very few studies on currency, gold, and crude oil, suggesting that investors and researchers focus more on stock markets than on the other assets in terms of the day-of-the-week and holiday effects.
We choose related studies on the basis of whether they use a time-varying volatility model to examine the asymmetry, the day-of-the-week, or holiday effects.

Table \ref{tab6} and \ref{tab7} exhibit the results of the literature.
\citeauthor{IT06}'s (\citeyear{IT06}) approach is closest to that of our study; to examine the seasonal patterns in the US stock markets, the study incorporates the asymmetric, day-of-the-week, and holiday effects into the SV model.
The study observed positive effects in the post-holiday returns.
However, it does not consider the positive effect in volatility on Thursday and pre-holiday.
Moreover, contrary to our result, the study observed asymmetry effects.
With respect to stock markets, we find that only some of our findings are similar to those in the other literature.
Although the approach in \cite{HB07} is closer to our approach than other studies on currency markets, his reports are not consistent with our results.
In a recent study by \cite{SK18}, day-of-the-week effects are not observed in the returns of the primary currencies, suggesting that the examined currency markets may have become efficient.
Interestingly, according to \cite{BL06}, the day-of-the-week effects on the returns and volatility are absent for gold, unlike stock and currency.
Moreover, it is worthy of attention that the asymmetry effects are not observed only in the case of gold markets, which is consistent with our result.
When analyzing the variance of the assets, \cite{AH16} examines how Bitcoin returns behave when compared to gold and the dollar-euro exchange rate.
Owing to their sensitivity to the federal funds rate and the medium of exchange characteristics, it is reported that Bitcoin returns behave identical to that of the exchange rate; Bitcoin returns also behave similar to that of gold, as reflected through Bitcoin's response to exchange rates and symmetrical reaction to good and bad news.
This suggests that Bitcoin shares a relationship with currency and gold.

The discussion yields the following findings:
(1) Bitcoin is associated with stock in relation to the day-of-the-week and holiday effects.
(2) Bitcoin is related to currency because of its reaction to the federal funds and the medium of exchange characteristics.
(3) Bitcoin reflects the characteristics of gold with respect to the absence of the asymmetry effect.
Therefore, we find that Bitcoin is positioned somewhere between stock, currency, and gold.

\section{Conclusions}\label{sec:conc}
Using the ASV model, this study investigates the day-of-the-week and holiday effects on the returns and volatility of Bitcoin from January 1, 2013 to August 31, 2019. 
National holidays and religious holidays in Japan, China, Germany, and the United States are used to verify the pre-holiday, holiday, and post-holiday effects.
The empirical results with respect to the day-of-the-week effects show that Friday and Thursday record larger returns and higher volatility, respectively, while smaller volatilities are recorded on Saturday and Sunday.
In terms of holiday effects, the positive post-holiday effect on returns and weak positive pre-holiday effect on volatility are only found in the United States.
An asymmetry effect is not observed in this study, supporting the absence of the asymmetry effect in the Bitcoin market.

This study can contribute to the study of Bitcoin volatility and risk control for investors.
First, the volatilities are small on Saturday and Sunday.
This might be related to the fact that the other markets like stock markets are closed on Saturday and Sunday.
Second, the holiday effect can be only observed in the United States.
This might be because the United States is the most active trading country.
These results suggests that the investors should pay attention to the trends on weekend and the holidays in the United States.
These are still potential implication, and further analysis is required.
However, concerning Bitcoin's positioning as a financial asset, the findings suggest that Bitcoin has a relationship with stock, currency, and gold with respect to the day-of-the-week and holiday effects, the reaction to the federal funds and the medium of exchange characteristics, and the lack of asymmetry effect, respectively.
Therefore, Bitcoin is assumed to be positioned somewhere between stock, currency, and gold.

This study provides an in-depth understanding to investors pursuing more profit and researchers undertaking empirical analyses of the Bitcoin market.
However, we could not explain the reason behind the occurrence of the day-of-the-week and holiday effects on returns and volatility in the Bitcoin market because it is difficult that Bitcoin is applied to existing theories.
This suggests that Bitcoin is the fourth unprecedented market, unlike other financial markets.
Therefore, further studies are needed to uncover the characteristics of Bitcoin; in this regard, future studies must also focus on developing a theory.

\bibliographystyle{ecta}
\bibliography{references}

\begin{thebibliography}{48}
\newcommand{\enquote}[1]{``#1''}
\expandafter\ifx\csname natexlab\endcsname\relax\def\natexlab#1{#1}\fi

\bibitem[\protect\citeauthoryear{Aggarwl}{Aggarwl}{2019}]{DA19}
\textsc{Aggarwl, D.} (2019): \enquote{Do {Bitcoins} Follow a Random Walk
  Model?} \emph{Research in Economics}, 73, 15--22.

\bibitem[\protect\citeauthoryear{Aharon and Qadan}{Aharon and
  Qadan}{2019}]{DY18}
\textsc{Aharon, D.~Y. and M.~Qadan} (2019): \enquote{Bitcoin and the
  Day-of-the-Week Effect,} \emph{Finance Research Letters}, 31.

\bibitem[\protect\citeauthoryear{Asai and McAleer}{Asai and
  McAleer}{2007}]{MM07}
\textsc{Asai, M. and M.~McAleer} (2007): \enquote{Dynamic Asymmetric Leverage
  in Stochastic Volatility Models,} \emph{Econometric Reviews}, 24, 317--332.

\bibitem[\protect\citeauthoryear{Baker, Rahman, and Saadi}{Baker
  et~al.}{2008}]{KB08}
\textsc{Baker, H.~K., A.~Rahman, and S.~Saadi} (2008): \enquote{The
  Day-of-the-Week Effect and Conditional Volatility: {Sensitivity} of Error
  Distributional Assumptions,} \emph{Review of Financial Economics}, 17,
  280--295.

\bibitem[\protect\citeauthoryear{Balaban, Bayar, and Kan}{Balaban
  et~al.}{2001}]{BBK01}
\textsc{Balaban, E., A.~Bayar, and {\"{O}}.~B. Kan} (2001): \enquote{Stock
  Returns, Seasonality and Asymmetric Conditional Volatility in World Equity
  Markets,} \emph{Applied Economics Letters}, 8, 263--268.

\bibitem[\protect\citeauthoryear{Baur, Cahill, Godfrey, and Liu}{Baur
  et~al.}{2019}]{DB19}
\textsc{Baur, D.~G., D.~Cahill, K.~Godfrey, and Z.~F. Liu} (2019):
  \enquote{Bitcoin Time-of-Day, Day-of-Week and Month-of-Year Effects in
  Returns and Trading Volume,} \emph{Finance Research Letters}, 31, 78--92.

\bibitem[\protect\citeauthoryear{Berument, Cosukun, and Sahin}{Berument
  et~al.}{2007}]{HB07}
\textsc{Berument, H., M.~N. Cosukun, and A.~Sahin} (2007): \enquote{Day of the
  Week Effect on Foreign Exchange Market Volatility: {Evidence} from {Turkey},}
  \emph{Research in International Business and Finance}, 21, 87--97.

\bibitem[\protect\citeauthoryear{Bollerslev}{Bollerslev}{1986}]{TB86}
\textsc{Bollerslev, T.} (1986): \enquote{Generalized Autoregressive Conditional
  Heteroskedasticity,} \emph{Journal of Econometrics}, 31, 307--327.

\bibitem[\protect\citeauthoryear{Bouoiyour, Selmii, Tiwari, and
  Olayeni}{Bouoiyour et~al.}{2016}]{JB16}
\textsc{Bouoiyour, J., R.~Selmii, A.~Tiwari, and O.~Olayeni} (2016):
  \enquote{What Drives {Bitcoin} Price?} \emph{Economics Bulletin}, 36,
  843--850.

\bibitem[\protect\citeauthoryear{Caporale and Plastun}{Caporale and
  Plastun}{2019}]{GM19}
\textsc{Caporale, G.~M. and A.~Plastun} (2019): \enquote{The Day of the Week
  Effect in the Cryptocurrency Market,} \emph{Finance Research Letters}, 31.

\bibitem[\protect\citeauthoryear{Casado, Muga, and Santamaria}{Casado
  et~al.}{2013}]{JC13}
\textsc{Casado, J., L.~Muga, and R.~Santamaria} (2013): \enquote{The Effect of
  {US} Holidays on the {European} Markets: {When} the cat's away...}
  \emph{Accounting \& Finance}, 53, 111--136.

\bibitem[\protect\citeauthoryear{Chib}{Chib}{2001}]{SC01}
\textsc{Chib, S.} (2001): \enquote{{Markov} Chain {Monte} {Carlo} Methods:
  {Computation} and Inference,} Elsevier, vol.~5 of \emph{Handbook of
  Econometrics}, 3569--3649.

\bibitem[\protect\citeauthoryear{Dodd and Gakhovich}{Dodd and
  Gakhovich}{2011}]{OD12}
\textsc{Dodd, O. and A.~Gakhovich} (2011): \enquote{The Holiday Effect in
  {Central} and {Eastern} {European} Financial Markets,} \emph{Investment
  Management and Financial Innovations}, 8, 29--35.

\bibitem[\protect\citeauthoryear{Doornik}{Doornik}{2013}]{D13}
\textsc{Doornik, J.~A.} (2013): \emph{Ox${}^{\text{TM}}$ 7: {An}
  Object-Oriented Matrix Programming Language}, London: Timberlake Consultants
  Press.

\bibitem[\protect\citeauthoryear{Dyhrberg}{Dyhrberg}{2016{\natexlab{a}}}]{AH16}
\textsc{Dyhrberg, A.~H.} (2016{\natexlab{a}}): \enquote{Bitcoin, gold and the
  dollar -- {A} {GARCH} volatility analysis,} \emph{Finance Research Letters},
  16, 85--92.

\bibitem[\protect\citeauthoryear{Dyhrberg}{Dyhrberg}{2016{\natexlab{b}}}]{AHD16}
---\hspace{-.1pt}---\hspace{-.1pt}--- (2016{\natexlab{b}}): \enquote{Hedging
  Capabilities of {Bitcoin}. {Is} It the Virtual Gold?} \emph{Finance Research
  Letters}, 16, 139--144.

\bibitem[\protect\citeauthoryear{Fama}{Fama}{1965}]{EF65}
\textsc{Fama, E.~F.} (1965): \enquote{The Behavior of Stock-Market Prices,}
  \emph{The Journal of Business}, 38, 34--105.

\bibitem[\protect\citeauthoryear{Fatemi and Park}{Fatemi and Park}{1996}]{FP96}
\textsc{Fatemi, A.~M. and J.~Park} (1996): \enquote{Seasonal Patterns in
  {Japanese} {ADR} Returns and the {US} Stock Market Influence,} \emph{Japan \&
  the World Economy}, 8, 65--79.

\bibitem[\protect\citeauthoryear{French}{French}{1980}]{KF80}
\textsc{French, K.~R.} (1980): \enquote{Stock Returns and the Weekend Effect,}
  \emph{Journal of Financial Economics}, 8, 55--69.

\bibitem[\protect\citeauthoryear{Gelfand and Smith}{Gelfand and
  Smith}{1990}]{GS90}
\textsc{Gelfand, A.~E. and A.~F.~M. Smith} (1990): \enquote{Sampling-Based
  Approaches to Calculating Marginal Densities,} \emph{Journal of the American
  Statistical Association}, 85, 398--409.

\bibitem[\protect\citeauthoryear{Geweke}{Geweke}{1992}]{GJ92}
\textsc{Geweke, J.} (1992): \enquote{Evaluating the Accuracy of Sampling-Based
  Approaches to the Calculation of Posterior Moments,} Oxford University Press,
  vol.~4 of \emph{Bayesian Statistics}, 168--182.

\bibitem[\protect\citeauthoryear{Holloway, Shankar, and Rahmanb}{Holloway
  et~al.}{2002}]{GH02}
\textsc{Holloway, G., B.~Shankar, and S.~Rahmanb} (2002): \enquote{Bayesian
  Spatial Probit Estimation: {A} Primer and an Application to {HYV} Rice
  Adoption,} \emph{Agricultural Economics}, 27, 383--402.

\bibitem[\protect\citeauthoryear{Hsieh}{Hsieh}{1988}]{DH88}
\textsc{Hsieh, D.~A.} (1988): \enquote{The Statistical Properties of Daily
  Foreign Exchange Rates: 1974--1983,} \emph{Journal of International
  Economics}, 24, 129--145.

\bibitem[\protect\citeauthoryear{Islam and Sultana}{Islam and
  Sultana}{2015}]{RI15}
\textsc{Islam, R. and N.~Sultana} (2015): \enquote{Day of the Week Effect on
  Stock Return and Volatility: {Evidence} from Chittagong Stock Exchange,}
  \emph{European Journal of Business and Management}, 7, 165--172.

\bibitem[\protect\citeauthoryear{Jiang, Nie, and Ruan}{Jiang
  et~al.}{2018}]{JY18}
\textsc{Jiang, Y., H.~Nie, and W.~Ruan} (2018): \enquote{Time-Varying Long-Term
  Memory in {Bitcoin} Market,} \emph{Finance Research Letters}, 25, 280--284.

\bibitem[\protect\citeauthoryear{Kim, Shephard, and Chib}{Kim
  et~al.}{1998}]{SK98}
\textsc{Kim, S., N.~Shephard, and S.~Chib} (1998): \enquote{Stochastic
  Volatility: {Likelihood} Inference and Comparison with {ARCH} Models,}
  \emph{The Review of Economic Studies}, 65, 361--393.

\bibitem[\protect\citeauthoryear{Kumar}{Kumar}{2018}]{SK18}
\textsc{Kumar, S.} (2018): \enquote{On the Disappearance of Calendar Anomalies:
  {Have} the Currency Markets Become Efficient?} \emph{Studies in Economics and
  Finance}, 35, 441--456.

\bibitem[\protect\citeauthoryear{Lucey and Tully}{Lucey and Tully}{2006}]{BL06}
\textsc{Lucey, B.~M. and E.~Tully} (2006): \enquote{Seasonality, Risk and
  Return in Daily {COMEX} Gold and Silver Data 1982--2002,} \emph{Applied
  Financial Economics}, 16, 319--333.

\bibitem[\protect\citeauthoryear{Ma and Tanizaki}{Ma and
  Tanizaki}{2019{\natexlab{a}}}]{DM19a}
\textsc{Ma, D. and H.~Tanizaki} (2019{\natexlab{a}}): \enquote{The
  Day-of-the-Week Effect on {Bitcoin} Return and Volatility,} \emph{Research in
  International Business and Finance}, 49, 127--136.

\bibitem[\protect\citeauthoryear{Ma and Tanizaki}{Ma and
  Tanizaki}{2019{\natexlab{b}}}]{DM19b}
---\hspace{-.1pt}---\hspace{-.1pt}--- (2019{\natexlab{b}}): \enquote{On the
  Day-of-the-Week Effects of {Bitcoin} Markets: {International} Evidence,}
  \emph{China Finance Review International}, 31, 455--478.

\bibitem[\protect\citeauthoryear{McKenzie}{McKenzie}{2002}]{MM02}
\textsc{McKenzie, M.} (2002): \enquote{The Economics of Exchange Rate
  Volatility Asymmetry,} \emph{International Journal of Financial \&
  Economics}, 7, 247--260.

\bibitem[\protect\citeauthoryear{Mitra}{Mitra}{2016}]{PM16}
\textsc{Mitra, P.} (2016): \enquote{Day-of-the-Week Effect on Stock Market
  Return and Volatility: {Evidence} from {Indian} Stock Market,} \emph{IOSR
  Journal of Economics and Finance}, 7, 99--107.

\bibitem[\protect\citeauthoryear{Nakajima}{Nakajima}{2012}]{JN12}
\textsc{Nakajima, J.} (2012): \enquote{Bayesian Analysis of Generalized
  Autoregressive Conditional Heteroskedasticity and Stochastic Volatility:
  {Modeling} Leverage, Jumps and Heavy-Tails for Financial Time Series,}
  \emph{The Japanese Economic Review}, 63, 81--103.

\bibitem[\protect\citeauthoryear{Nakajima and Omori}{Nakajima and
  Omori}{2009}]{JN09}
\textsc{Nakajima, J. and Y.~Omori} (2009): \enquote{Leverage, Heavy-Tails and
  Correlated Jumps in Stochastic Volatility Models,} \emph{Computational
  Statistics \& Data Analysis}, 53, 2335--2353, the Fourth Special Issue on
  Computational Econometrics.

\bibitem[\protect\citeauthoryear{Nakamoto}{Nakamoto}{2008}]{SN08}
\textsc{Nakamoto, S.} (2008): \enquote{Bitcoin: A peer-to-peer electronic cash
  system,} .

\bibitem[\protect\citeauthoryear{Nishino and Kakamu}{Nishino and
  Kakamu}{2015}]{NK15}
\textsc{Nishino, H. and K.~Kakamu} (2015): \enquote{A Random Walk Stochastic
  Volatility Model for Income Inequality,} \emph{Japan {\&} the World Economy},
  36, 21--28.

\bibitem[\protect\citeauthoryear{Nishino, Kakamu, and Oga}{Nishino
  et~al.}{2012}]{NKO12}
\textsc{Nishino, H., K.~Kakamu, and T.~Oga} (2012): \enquote{Bayesian
  Estimation of Persistent Income Inequality Using the Lognormal Stochastic
  Volatility Model,} \emph{Journal of Income Distribution}, 21, 88--101.

\bibitem[\protect\citeauthoryear{Omori, Chib, Shephard, and Nakajima}{Omori
  et~al.}{2007}]{YO07}
\textsc{Omori, Y., S.~Chib, N.~Shephard, and J.~Nakajima} (2007):
  \enquote{Stochastic Volatility with Leverage: {Fast} and Efficient Likelihood
  Inference,} \emph{Journal of Econometrics}, 140, 425--449.

\bibitem[\protect\citeauthoryear{Omori and Watanabe}{Omori and
  Watanabe}{2008}]{YO08}
\textsc{Omori, Y. and T.~Watanabe} (2008): \enquote{Block Sampler and Posterior
  Mode Estimation for Asymmetric Stochastic Volatility Models,}
  \emph{Computational Statistics {\&} Data Analysis}, 52, 2892--2910.

\bibitem[\protect\citeauthoryear{{R Core Team}}{{R Core Team}}{2020}]{R19}
\textsc{{R Core Team}} (2020): \emph{R: A Language and Environment for
  Statistical Computing}, R Foundation for Statistical Computing, Vienna,
  Austria.

\bibitem[\protect\citeauthoryear{Tanizaki}{Tanizaki}{2004}]{HT04}
\textsc{Tanizaki, H.} (2004): \enquote{On Asymmetry, Holiday and
  Day-of-the-Week Effects in Volatility of Daily Stock Returns: {The} Case of
  {Japan},} \emph{Journal of the Japan Statistical Society}, 34, 129--152.

\bibitem[\protect\citeauthoryear{Tiwari, Kumar, and Pathak}{Tiwari
  et~al.}{2019}]{AK19}
\textsc{Tiwari, A.~K., S.~Kumar, and R.~Pathak} (2019): \enquote{Modeling the
  Dynamics of {Bitcoin} and {Litecoin}: {GARCH} Versus Stochastic Volatility
  Models,} \emph{Applied Economics}, 51, 4073--4082.

\bibitem[\protect\citeauthoryear{Tsiakas}{Tsiakas}{2006}]{IT06}
\textsc{Tsiakas, I.} (2006): \enquote{Periodic Stochastic Volatility and Fat
  Tails,} \emph{Journal of Financial Econometrics}, 4, 90--135.

\bibitem[\protect\citeauthoryear{Tsiakas}{Tsiakas}{2010}]{IT10}
---\hspace{-.1pt}---\hspace{-.1pt}--- (2010): \enquote{The Economic Gains of
  Trading Stocks Around Holidays,} \emph{Journal of Financial Research}, 33,
  1--26.

\bibitem[\protect\citeauthoryear{Tully and Lucey}{Tully and Lucey}{2007}]{ET07}
\textsc{Tully, E. and B.~M. Lucey} (2007): \enquote{A Power {GARCH} Examination
  of the Gold Market,} \emph{Research in International Business and Finance},
  21, 316--325.

\bibitem[\protect\citeauthoryear{Yaya, Ogbonna, and Olubusoye}{Yaya
  et~al.}{2019}]{OY19}
\textsc{Yaya, O.~S., A.~E. Ogbonna, and O.~E. Olubusoye} (2019): \enquote{How
  Persistent and Dynamic Inter-Dependent Are Pricing of {Bitcoin} to Other
  Cryptocurrencies Before and After 2017/18 Crash?} \emph{Physica A:
  Statistical Mechanics and its Applications}, 531, 121732.

\bibitem[\protect\citeauthoryear{Yu}{Yu}{2005}]{JY05}
\textsc{Yu, J.} (2005): \enquote{On Leverage in a Stochastic Volatility Model,}
  \emph{Journal of Econometrics}, 127, 165--178.

\bibitem[\protect\citeauthoryear{Yuan and Gupta}{Yuan and Gupta}{2014}]{TR14}
\textsc{Yuan, T. and R.~Gupta} (2014): \enquote{Chinese {Lunar} {New} {Year}
  Effect in {Asia} Stock Markets, 1999--2012,} \emph{The Quarterly Review of
  Economics and Finance}, 54, 529--537.

\end{thebibliography}

\clearpage

\clearpage
\renewcommand{\baselinestretch}{1.0}
\begin{table}[tbh]
	\begin{center}
		\caption{Descriptive statistics for returns on day-of-the weeks}
		\begin{tabular*}{130mm}{@{\extracolsep{\fill}}lrrrrrrr}
			\hline
			& \multicolumn{1}{c}{Obs} & \multicolumn{1}{c}{Mean} & \multicolumn{1}{c}{SD} & \multicolumn{1}{c}{Min} & \multicolumn{1}{c}{Max} & \multicolumn{1}{c}{Skew} & \multicolumn{1}{c}{Kurt}\\
			\hline
			All       & 2434 &  0.270 & 4.742 & -66.495 & 30.747 & -1.231 & 25.510\\
			Sunday    &  347 &  0.116 & 3.573 & -14.886 & 12.943 & -0.150 &  5.766\\
			Monday    &  347 &  0.715 & 4.568 & -24.265 & 30.403 &  0.435 & 11.937\\
			Tuesday   &  347 &  0.413 & 4.699 & -18.464 & 21.341 & -0.371 &  7.390\\
			Wednesday &  348 & -0.015 & 5.426 & -35.844 & 30.747 & -0.994 & 13.950\\
			Thursday  &  348 &  0.083 & 6.205 & -66.495 & 26.465 & -3.161 & 41.828\\
			Friday    &  348 &  0.314 & 4.457 & -21.992 & 29.616 &  0.267 & 10.876\\
			Saturday  &  348 &  0.263 & 3.674 & -15.846 & 15.529 & -0.227 &  6.902\\
			\hline
		\end{tabular*}
		\label{tab1}
	\end{center}
\end{table}

\begin{table}[tbh]
	\begin{center}
		\caption{Descriptive statistics for returns on pre-holidays, holidays and post-holidays in each country }
		\begin{tabular*}{130mm}{@{\extracolsep{\fill}}llrrrrrrr}
			\hline
			& & \multicolumn{1}{c}{Obs} & \multicolumn{1}{c}{Mean} & \multicolumn{1}{c}{SD} & \multicolumn{1}{c}{Min} & \multicolumn{1}{c}{Max} & \multicolumn{1}{c}{Skew} & \multicolumn{1}{c}{Kurt}\\
			\hline
			\multicolumn{2}{l}{Pre-holiday}&&&&&&&\\		
			& Japan             &  83 &  0.144 & 3.924 & -11.809 & 14.699 &  0.148 &  6.112\\
			& China             &  45 &  0.718 & 4.208 & -10.886 & 11.870 & -0.068 &  4.775\\
			& Germany           &  54 & -0.640 & 4.054 & -20.331 &  7.244 & -2.176 & 11.855\\
			& the United States &  65 &  0.041 & 4.336 & -12.251 & 10.450 & -0.325 &  3.893\\
			\multicolumn{2}{l}{Holiday}&&&&&&&\\
			& Japan             & 139 &  1.159 & 3.783 & -10.080 & 15.529 &  0.619 &  5.635\\
			& China             & 176 &  0.432 & 3.584 & -20.331 & 12.168 & -1.525 & 13.203\\
			& Germany           &  61 &  0.311 & 4.505 & -18.084 & 12.168 & -0.466 &  7.126\\
			& the United States &  72 &  0.276 & 3.186 &  -8.946 &  7.971 & -0.439 &  3.854\\
			\multicolumn{2}{l}{Post-holiday}&&&&&&&\\
			& Japan             &  84 & -0.107 & 4.353 & -18.116 & 10.753 & -1.115 & 6.989\\
			& China             &  46 &  0.238 & 4.537 & -18.084 & 12.943 & -0.924 & 8.272\\
			& Germany           &  55 &  0.605 & 3.871 &  -8.792 & 11.909 &  0.143 & 4.080\\
			& the United States &  65 &  1.197 & 5.130 & -17.313 & 13.880 & -0.783 & 6.381\\	 
			\hline
		\end{tabular*}
		\label{tab2}
	\end{center}
\end{table}

\begin{table}[tbh]
	\begin{center}
		\caption{Estimation results of the observation equation in the ASV model}
		\begin{tabular*}{120mm}{@{\extracolsep{\fill}}llrrrrr}
			\hline
			\multicolumn{2}{l}{Parameter}	 & \multicolumn{1}{c}{Mean} & \multicolumn{1}{c}{SD} & \multicolumn{1}{c}{95\%CI} & \multicolumn{1}{c}{CD} & \multicolumn{1}{c}{IF}\\
			\hline	
			&  Constant  & -0.021 & 0.113 & [-0.245, 0.201] & 0.331 & 2.058 \\
			\multicolumn{2}{l}{Day-of-the-week}&&&&&\\
			& Sunday            &         0.076  & 0.149 & [-0.218,\ 0.371]  & 0.452 & 1.495\\
			& Monday            &         0.063  & 0.167 & [-0.261,\ 0.389]  & 0.861 & 4.283\\
			& Tuesday           &         0.303  & 0.172 & [-0.031,\ 0.641]  & 0.411 & 3.795\\
			& Thursday          &         0.161  & 0.170 & [-0.172,\ 0.491]  & 0.560 & 1.979\\
			& Friday            & \textbf{0.415} & 0.158 & [\ 0.107,\ 0.728] & 0.866 & 2.966\\
			& Saturday          &         0.216  & 0.146 & [-0.073,\ 0.504]  & 0.608 & 1.793\\
			\multicolumn{2}{l}{Pre-holiday}&&&&\\
			& Japan             &        -0.106  & 0.205 & [-0.507,\ 0.300]  & 0.205 & 2.333\\
			& China             &         0.104  & 0.259 & [-0.431,\ 0.598]  & 0.287 & 1.380\\
			& Germany           &        -0.388  & 0.295 & [-0.946,\ 0.209]  & 0.334 & 3.712\\
			& the United States &         0.302  & 0.363 & [-0.402,\ 1.018]  & 0.135 & 1.573\\
			\multicolumn{2}{l}{Holiday}&&&&&\\
			& Japan             &         0.051  & 0.190 & [-0.319,\ 0.430]  & 0.079 & 3.531\\
			& China             &         0.092  & 0.130 & [-0.163,\ 0.349]  & 0.902 & 3.092\\
			& Germany           &        -0.051  & 0.278 & [-0.603,\ 0.491]  & 0.859 & 1.636\\
			& the United States &        -0.044  & 0.286 & [-0.603,\ 0.508]  & 0.421 & 3.045\\
			\multicolumn{2}{l}{Post-holiday}&&&&&\\
			& Japan             &         0.287  & 0.244 & [-0.186,\ 0.767]  & 0.836 & 3.293\\
			& China             &         0.322  & 0.256 & [-0.182,\ 0.830]  & 0.520 & 1.850\\
			& Germany           &        -0.288  & 0.286 & [-0.851,\ 0.281]  & 0.439 & 2.663\\
			& the United States & \textbf{0.784} & 0.402 & [\ 0.015,\ 1.606] & 0.346 & 2.851\\
			\hline	
		\end{tabular*}		
		\label{tab3}

		\begin{minipage}{120mm}
			The posterior means, which are indicated in bold, do not include zero in the 95\% credible interval.
		\end{minipage}
	\end{center}
\end{table}	

\begin{table}[tbh]
	\begin{center}
		\caption{Estimation results of the state-space equation in the ASV model}
		\begin{tabular*}{120mm}{@{\extracolsep{\fill}}llrrrrr}
			\hline
			\multicolumn{2}{l}{Parameter}	 & \multicolumn{1}{c}{Mean} & \multicolumn{1}{c}{SD} & \multicolumn{1}{c}{95\%CI} & \multicolumn{1}{c}{CD} & \multicolumn{1}{c}{IF}\\
			\hline	
			& Constant      &  \textbf{2.171} & 0.173 & [\ 1.830,\ \ 2.511] & 0.284 &  4.561\\	
			& $\phi$        &  \textbf{0.918} & 0.013 & [\ 0.890,\ \ 0.942] & 0.425 & 37.105\\
			& $\sigma$      &  \textbf{0.565} & 0.044 & [\ 0.483,\ \ 0.656] & 0.246 & 64.279\\
			& $\rho$        &          0.048  & 0.048 & [-0.047,\ \ 0.141]  & 0.670 &  9.875\\
			\multicolumn{2}{l}{Day-of-the-week} & & & & & \\
			& Sunday        & \textbf{-0.394} & 0.135 & [-0.660,\ -0.126]   & 0.069 & 18.856\\
			& Monday        &          0.054  & 0.134 & [-0.216,\ \ 0.311]  & 0.679 & 13.042\\
			& Tuesday       &          0.096  & 0.135 & [-0.169,\ \ 0.363]  & 0.289 &  9.882\\
			& Thursday      &  \textbf{0.289} & 0.132 & [\ 0.028,\ \ 0.544] & 0.100 & 20.574\\
			& Friday        &         -0.044  & 0.134 & [-0.309,\ \ 0.218]  & 0.265 &  9.271\\
			& Saturday      & \textbf{-0.332} & 0.134 & [-0.598,\ -0.074]   & 0.578 & 11.127\\
			\multicolumn{2}{l}{Pre-holiday}&&&&&\\
			& Japan         &         -0.341  & 0.208 & [-0.746,\ \ 0.071]  & 0.745 & 20.810\\
			& China         &         -0.039  & 0.294 & [-0.598,\ \ 0.550]  & 0.093 &  7.749\\
			& Germany       &          0.169  & 0.285 & [-0.391,\ \ 0.727]  & 0.904 & 10.331\\
			& United States &          0.462  & 0.246 & [-0.006,\ \ 0.964]  & 0.858 &  8.897\\
			\multicolumn{2}{l}{Holiday}&&&&&\\
			& Japan         &         -0.203  & 0.180 & [-0.554,\ \ 0.152]  & 0.351 &  6.546\\
			& China         &         -0.325  & 0.190 & [-0.690,\ \ 0.055]  & 0.937 &  4.892\\
			& Germany       &          0.138  & 0.262 & [-0.365,\ \ 0.656]  & 0.230 &  7.913\\
			& United States &         -0.350  & 0.240 & [-0.817,\ \ 0.120]  & 0.160 &  8.639\\
			\multicolumn{2}{l}{Post-holiday}&&&&&\\
			& Japan         &         -0.176  & 0.207 & [-0.575,\ \ 0.230]  & 0.338 &  6.802\\
			& China         &          0.015  & 0.279 & [-0.513,\ \ 0.578]  & 0.770 & 11.897\\
			& Germany       &          0.114  & 0.263 & [-0.396,\ \ 0.644]  & 0.092 & 13.620\\
			& United States &          0.274  & 0.251 & [-0.217,\ \ 0.767]  & 0.728 &  4.312\\
			\hline
		\end{tabular*}
		\label{tab4}	

		\begin{minipage}{120mm}
			The posterior means, which are indicated in bold, do not include zero in the 95\% credible interval.
		\end{minipage}
	\end{center}
\end{table}

\begin{table}[t]
	\centering
	\caption{The summarization of literature for stock, currency, and gold}
	\begin{tabular*}{140mm}{@{\extracolsep{\fill}}lccccccc}
		\hline
		\multicolumn{4}{c}{} & \multicolumn{2}{c}{DOW${}^{1}$} & \multicolumn{2}{c}{HOL${}^{2}$}\\
		\cline{5-6}\cline{7-8}
		\multicolumn{1}{c}{Author} & Type of assets & Model & Asy${}^{3}$ & R & V & R & V\\
		\hline	
		% The Statistical Properties of Daily Foreign Exchange Rates: 1974-1983 
		\cite{DH88} & Currency & ARCH & - & \circle & \circle & \circle & \circle\\
		\hline
		% Stock Returns, Seasonality and Asymmetric Conditional Volatility in World Equity Markets
		\cite{BBK01} & Stock & GARCH & \circle & \circle & \circle & - & - \\
		\hline
		% On Asymmetry, Holiday and Day-of-the-week Effects in Volatility of Daily Stock Returns: The Case of Japan
		\cite{HT04} & Stock & SV & \circle & - & \circle & - & \circle\\
		\hline	
		\cite{BL06} & Gold & GARCH & \circle & \circle & \circle & - & - \\
		\hline
		% Periodic Stochastic Volatility and Fat Tails
		\cite{IT06} & Stock & SV & \circle & \circle & \circle & \circle & \circle\\
		\hline
		% Day-of-the-week effect on foreign exchange market volatility: Evidence from Turkey
		\cite{HB07} & Currency & EGARCH & \circle & \circle & \circle & - & - \\
		\hline
		% A power GARCH examination of the gold market
		\cite{ET07} & Gold & APGARCH & \circle & - & - & - & - \\
		\hline
		% The day-of-the-week effect and conditional volatility: Sensitivity of error distributional assumptions
		\cite{KB08} & Stock & GARCH & - & \circle & \circle & - & - \\
		\hline
		% The economic gains of trading stocks around holidays
		\cite{IT10} & Stock & SV & \circle & - & - & \circle & \circle\\
		\hline
		% The effect of US holidays on the European markets: when the cat\UTF{0081}fs away\UTF{0081}c
		\cite{JC13} &	Stock & GJR-GARCH-M & - & \circle & - & \circle & - \\
		\hline		
		% Day-of-the-Week Effect on Stock Market Return and Volatility: Evidence from Indian Stock Market
		\cite{PM16} & Stock & GARCH & - & \circle & \circle & - & - \\
		\hline
		% On the disappearance of calendar anomalies: have the currency markets become efficient?
		\cite{SK18} & Currency & GARCH & - & \circle & - & - & - \\
		\hline
	\end{tabular*}
	\label{tab5}
	
	\begin{minipage}{120mm}
		\circle: examined \ - : not examined\\
		${}^{1}$ Day-of-the-week effects\\
		${}^{2}$ Holiday effects\\
		${}^{3}$ Asymmetry effect
	\end{minipage}
\end{table}

\begin{landscape}
	\begin{table}[t]
		\centering
		\caption{The results of literature for stock, currency, and gold in part I}
		\begin{tabular*}{240mm}{@{\extracolsep{\fill}}llcccccccccccccccccc}
			\hline
			\multicolumn{4}{c}{} & \multicolumn{10}{c}{DOW} & \multicolumn{6}{c}{HOL}\\
			\cline{5-14}\cline{15-20}
			\multicolumn{4}{c}{} & \multicolumn{2}{c}{Mon} & \multicolumn{2}{c}{Tue} & \multicolumn{2}{c}{Wed} & \multicolumn{2}{c}{Thu} & \multicolumn{2}{c}{Fri} & \multicolumn{2}{c}{Pre-} & \multicolumn{2}{c}{Post-} & \multicolumn{2}{c}{Length${}^{1}$}\\
			\cline{5-6}\cline{7-8}\cline{9-10}\cline{11-12}\cline{13-14}\cline{15-16}\cline{17-18}\cline{19-20}
			\multicolumn{1}{c}{Author} & \multicolumn{1}{c}{Period} & Used data & Asy & R & V & R & V & R & V & R & V & R & V & R & V & R & V & R & V\\
			\hline
			% The Statistical Properties of Daily Foreign Exchange Rates: 1974-1983 
			\cite{DH88} & 1/2, 1974 - 12/30, 1983 & GBP - USD${}^{2}$ & - & \cross & \plus & \cross & \plus & \plus & \plus & \cross & \plus & C & C & - & - & - & - & \cross & \plus\\ 
			{ } & { } & CAD - USD${}^{2}$ & - & \cross & \plus	 & \plus	& \cross & \cross & \cross & \plus & \plus & C & C & - & - & - & - & \cross & \plus\\ 
			{ } & { } & DEM - USD${}^{2}$ & - & \cross & \plus & \cross & \cross & \plus & \plus & \cross & \plus & C & C & - & - & - & - & \cross & \plus\\
			{ } & { } & JPY - USD${}^{2}$ & -	& \cross & \plus & \cross & \plus & \cross & \plus & \cross & \plus & C & C & - & - & - & - & \cross & \minus\\
			{ } & { } & CHF - USD${}^{2}$ & - & \cross & \plus & \cross & \plus & \plus & \plus & \cross & \plus & C & C & -& - & - & - & \cross & \plus\\
			\hline
			% Stock Returns, Seasonality and Asymmetric Conditional Volatility in World Equity Markets
			\cite{BBK01} & 7/20, 1993 - 7/1, 1998 & MSCI AUS${}^{3}$ & \cross & C & C & \cross & \cross & \cross & \cross & \cross & \cross & \cross & \cross & - & - & - & - & - & - \\
			{ } & { } & MSCI GER${}^{3}$ & \circle & C & C & \minus & \cross & \cross & \cross & \cross & \cross & \minus & \cross & - & - & - & - & - & - \\	
			{ } & { } & MSCI HON${}^{3}$ & \circle & C & C & \cross & \cross & \plus & \cross & \cross & \cross & \cross & \cross & - & - & - & - & - & - \\	
			{ } & { } & MSCI JAP${}^{3}$ & \cross & C & C & \plus & \cross & \plus & \cross & \plus & \cross & \cross & \cross & - & - & - & - & - & - \\
			{ } & { } & MSCI UK${}^{3}$ & \cross & C & C & \cross & \cross & \cross & \cross & \cross & \cross & \cross & \cross & - & - & - & - & - & - \\	
			{ } & { } & MSCI US${}^{3}$ & \circle & C & C & \cross & \plus & \cross & \cross & \cross & \plus & \cross & \cross & - & - & - & - & - & - \\
			\hline
			% On Asymmetry, Holiday and Day-of-the-week Effects in Volatility of Daily Stock Returns: The Case of Japan
			\cite{HT04} & 1/4, 1985 - 6/10, 2004 & Nikkei 225${}^{4}$ & \circle & - & \plus & - & \minus & - & \cross & - & \cross & - & \cross & - & - & - & - & - & \plus\\
			\hline
			\cite{BL06} & 4/1, 1982 - 11/28, 2002 & Cash gold${}^{5}$ & \triangle & \triangle & \triangle & - & \triangle & - & \triangle & - & \triangle & - & C & - & - & - & - & - \\
			\hline
			% Periodic Stochastic Volatility and Fat Tails
			\cite{IT06} & 7/2, 1962 - 12/31, 2003 & DJIA${}^{6}$ & \circle & \minus & \plus & \cross & \cross & \cross & \cross & \minus & \cross & C & C & \cross & \minus & \plus & \cross & - & - \\
			& & S{\&}P500${}^{7}$ & \circle & \minus & \plus & \minus & \plus & \cross & \cross & \minus & \cross & C & C & \plus & \minus & \plus & \cross & - & - \\
			\hline
			% Day-of-the-week effect on foreign exchange market volatility: Evidence from Turkey 
			\cite{HB07} & 3/12, 2001 - 11/22, 2005 & TL - USD  & \circle${}^{8}$ & \cross & \cross & \cross & \plus & C & C & \cross & \cross & \cross & \cross & - & - & - & - & - & - \\
			\hline		 					
	\end{tabular*}
	\label{tab6}

	\begin{minipage}{200mm}
		\circle: asymmetry effect \  \plus: positive effect \ \minus: negative effect \ \triangle: little effect \  \cross: no effect \ C: constant 
		\vspace{3mm}

		${}^{1}$ Create dummies by taking the number of holidays between $(t-1)$th and $t$th trading day.\\
		${}^{2}$ the British pound, Canadian dollar, Germany Mark, Japanese yen and Swiss franc against the USD, respectively.\\
		${}^{3}$ Stock indices against the USD in Australia, Germany, Hong Kong, Japan, the United Kingdom and the United States by the Morgan Stanley Capital International Index, respectively.\\
		${}^{4}$ Stock market index measured to determine the performance of the most liquid 225 stocks on the Tokyo Stock Exchange.\\
		${}^{5}$ Gold price against the USD from COMEX.\\
		${}^{6}$ Stock market index measured on the stock performance of the major 30 companies in the United States.\\
		${}^{7}$ Stock market index measured to determine the stock performance of the major 500 companies in the United States.\\
		${}^{8}$ The asymmetry effect is observed; currency markets are generally symmetric.
	\end{minipage}
 \end{table}
\end{landscape}
 
\begin{landscape}
	\begin{table}[t]
		\centering
		\caption{The results of literature for stock, currency, and gold in part II}
		\begin{tabular*}{240mm}{@{\extracolsep{\fill}}llcccccccccccccccccc}
			\hline
			\multicolumn{4}{c}{} & \multicolumn{10}{c}{DOW} & \multicolumn{6}{c}{HOL}\\
			\cline{5-14}\cline{15-20}
			\multicolumn{4}{c}{} & \multicolumn{2}{c}{Mon} & \multicolumn{2}{c}{Tue} & \multicolumn{2}{c}{Wed} & \multicolumn{2}{c}{Thu} & \multicolumn{2}{c}{Fri} & \multicolumn{2}{c}{Pre-} & \multicolumn{2}{c}{Post-} & \multicolumn{2}{c}{Length}\\
			\cline{5-6}\cline{7-8}\cline{9-10}\cline{11-12}\cline{13-14}\cline{15-16}\cline{17-18}\cline{19-20}
			\multicolumn{1}{c}{Author} & \multicolumn{1}{c}{Period} & Used data & Asy & R & V & R & V & R & V & R & V & R & V & R & V & R & V & R & V\\
			\hline
			% A power GARCH examination of the gold market
			\cite{ET07} & 1983 - 2003 & Gold Cash & \cross & - & - & - & - & - & - & - & - & - & - & - & - & - & - & - & - \\
			\hline
			% The day\UTF{0081}-of\UTF{0081}-the-week effect and conditional volatility: Sensitivity of error distributional assumptions
			\cite{KB08} & 1/3, 1977 - 3/31, 2002 & S\&P500/TSX${}^{1}$ & - & \minus & \minus & C & C & \cross & \minus & \cross & \minus & \plus & \minus & - & - & - & - & - & - \\
			\hline
			% The economic gains of trading stocks around holidays	
			\cite{IT10} & 7/2, 1962 - 12/30, 2005 & S{\&}P500 & \circle & - & - & - & - & - & - & - & - & - & - & \plus & \minus & \plus & \minus & - & - \\
			& & EW CRSP${}^{2}$ & \circle & - & - & - & - & - & - & - & - & - & - & \plus & \minus & \plus & \minus & - & - \\
			& & VW CRSP${}^{3}$ & \circle & - & - & - & - & - & - & - & - & - & - & \plus & \minus & \plus & \minus & - & - \\
			\hline
			% The effect of US holidays on the European markets: when the cat\UTF{0081}fs away\UTF{0081}c
			\cite{JC13} & 1991 - 2008 & DAX30${}^{4}$ & - & \cross & - & \cross & - & C & - & \cross & - & \cross & - & \plus & - & - & - & - & - \\
			& & FTSE100${}^{5}$ & - & \cross & - & \cross & - & C & - & \cross & - & \cross & - & \cross & - & - & - & - & - \\
			& & CAC40${}^{6}$ & - & \cross & - & \cross & - & C & - & \cross & - & \cross & - & \cross & - & - & - & - & - \\
			& & IBEX35${}^{7}$ & - & \cross & - & \cross & - & C & - & \cross & - & \cross & - & \cross & - & - & - & - & - \\
			\hline
			% Day-of-the-Week Effect on Stock Market Return and Volatility: Evidence from Indian Stock Market
			\cite{PM16} &
			1/2, 2000 - 12/31- 2015 & BSE Sensex${}^{8}$ & - & C & C & \cross & \minus & \cross & \cross & \cross & \cross & \cross & \cross & - & - & - & - & - & - \\
			& & NSE 50${}^{9}$ & - & C & C & \cross & \minus & \cross & \cross & \cross & \cross & \cross & \cross & - & - & - & - & - & - \\
			\hline
			% On the disappearance of calendar anomalies: have the currency markets become efficient?	
			\cite{SK18} & Jan., 1994 - Dec., 2004 & GBP - USD & - & \minus & - & \cross & - & \cross & - & \plus & - & \plus & - & - & - & - & - & - & - \\
			& & CAD - USD & - & \minus & - & \cross & - & \minus & - & \cross & - & \plus & - & - & - & - & - & - & - \\
			& & JPY - USD & - & \cross & - & \cross & - & \minus & - & \cross & - & \plus & - & - & - & - & - & - & - \\
			& Jan., 2005 - Dec., 2014 & GBP - USD & - & \cross & - & \cross & - & \cross & - & \cross & - & \cross & - & - & - & - & - & - & - \\
			& & CAD - USD  & - & \cross & - & \cross & - & \cross & - & \cross & - & \cross & - & - & - & - & - & - & - \\
			& & JPY - USD & - & \cross & - & \cross & - & \cross & - & \cross & - & \cross & - & - & - & - & - & - & - \\
			\hline
		\end{tabular*}
		\label{tab7}
		\begin{minipage}[]{200mm}
			${}^{1}$ Composite price index from the Toronto Stock Exchange.\\
			${}^{2}$ Stock index from the equally weighted Center for Research in Security Prices portfolio.\\
			${}^{3}$ Stock index from the value-weighted Center for Research in Security Prices portfolio.\\
			${}^{4}$ Stock index of the 30 major German  companies.\\
			${}^{5}$ Stock index of the 100 companies listed on the London Stock Exchange with the highest market capitalization.\\
			${}^{6}$ Stock index measured to determine the performance of the 40 most liquid French stocks among the 100 largest market caps.\\
			${}^{7}$ Stock index measured to determine the performance of the 35 most liquid Spanish stocks.\\
			${}^{8}$ Stock index measured to determine the performance of the 30 well-established and financially sound companies in India.\\		
			${}^{9}$ Stock index measured to determine the performance of 50 Indian companies.
		\end{minipage}
	\end{table}
\end{landscape}
\end{document}